\documentclass[prx,twocolumn,english,superscriptaddress,floatfix,longbibliography]{revtex4-1} 

\usepackage{float}
\usepackage{bbm}
\usepackage{epsfig}
\usepackage{epstopdf}
\usepackage{graphicx, subfigure}
\usepackage{pgfplots}
\usetikzlibrary{patterns}
\usepgfplotslibrary{groupplots}
\usepackage{amssymb}
\usepackage{amsmath}
\usepackage{scalefnt}
\usepackage{color}
\usepackage[title]{appendix}
\usepackage{hyperref}
\usepackage{qcircuit}
\usepackage{algorithm} 
\usepackage{algpseudocode}
\usepackage[capitalize]{cleveref}
\usepackage{braket,mleftright}
\usepackage{physics}
\usepackage{import}
\usepackage{dcolumn}
\usepackage{csquotes}
\usepackage{tikz}
\usetikzlibrary{calc}
\usepackage{comment}

\floatstyle{plaintop}
\restylefloat{table}
\usepgfplotslibrary{fillbetween}
\pgfplotsset{
    discard if not/.style 2 args={
        x filter/.code={
            \edef\tempa{\thisrow{#1}}
            \edef\tempb{#2}
            \ifx\tempa\tempb
            \else
                
            \fi
        }
    }
}

\definecolor{teal}{RGB}{9,164,179}
\definecolor{my_orange}{RGB}{255,139,56}

\def\addlegendimage{\csname pgfplots@addlegendimage\endcsname}

\pgfplotsset{compat=1.16}

\begin{document}

\title{An Adaptive Quantum Approximate Optimization Algorithm for Solving Combinatorial Problems on a Quantum Computer}
    
    \author{Linghua Zhu}
        \email{zlinghua18@vt.edu}
        \affiliation{Department of Physics, Virginia Tech, Blacksburg, VA 24061, U.S.A}
    \author{Ho Lun Tang}%
        \affiliation{Department of Physics, Virginia Tech, Blacksburg, VA 24061, U.S.A}
    \author{George S. Barron}%
        \affiliation{Department of Physics, Virginia Tech, Blacksburg, VA 24061, U.S.A}
   \author{F. A. Calderon-Vargas}%
        \affiliation{Department of Physics, Virginia Tech, Blacksburg, VA 24061, U.S.A}    
   \author{Nicholas J. Mayhall}%
        \affiliation{Department of Chemistry, Virginia Tech, Blacksburg, VA 24061, U.S.A}
         \author{Edwin Barnes}%
        \affiliation{Department of Physics, Virginia Tech, Blacksburg, VA 24061, U.S.A}
    \author{Sophia E. Economou}%
        \email{economou@vt.edu}
        \affiliation{Department of Physics, Virginia Tech, Blacksburg, VA 24061, U.S.A}
    
    \date{\today}
    
    \begin{abstract}
       The quantum approximate optimization algorithm (QAOA) is a hybrid variational quantum-classical algorithm that solves combinatorial optimization problems. While there is evidence suggesting that the fixed form of the standard QAOA ansatz is not optimal, there is no systematic approach for finding better ans\"atze. We address this problem by developing an iterative version of QAOA that is problem-tailored, and which can also be adapted to specific hardware constraints. We simulate the algorithm on a class of Max-Cut graph problems and show that it converges much faster than the standard QAOA, while simultaneously reducing the required number of CNOT gates and optimization parameters. We provide evidence that this speedup is connected to the concept of shortcuts to adiabaticity.
    \end{abstract}
    
    \maketitle
    \section{Introduction}\label{sec:introduction}

        Many important computationally hard combinatorial optimization problems such as Max-Cut, graph coloring, traveling salesman, and scheduling management~\cite{Oh2019, kochenberger2014, Lucas2014ising} can be mapped to Ising Hamiltonians whose ground states provide the solutions. One can in principle solve these optimization problems on a quantum computer by initializing the quantum device in the ground state of a Hamiltonian that is easy to prepare and adiabatically tuning the latter into the problem Hamiltonian. In a digital quantum computer, this translates into a trotterized version of the adiabatic evolution operator, which is the alternating product of the evolution operators corresponding to the initial mixer and the problem (Ising) Hamiltonians. In the limit of an infinite product, this trotterized form becomes exact.
        
        QAOA is a hybrid quantum-classical variational algorithm that uses a finite-order version of this evolution operator to prepare wavefunction ans\"atze on a quantum processor~\cite{Farhi2014,Farhi2014design,Farhi2016quantum,Shaydulin2019}. QAOA is performed by variationally minimizing the expectation value of the Ising Hamiltonian with respect to the parameters in the ansatz. The quantum processor is also used to measure energy expectation values, while the optimization is done on a classical computer. There has been a lot of progress on QAOA recently on both the experimental and theoretical fronts~\cite{Pagano2019,Oh2019, Wang2018PRA,Zhou2018QAOA, Crooks2018, Yang2017PRX, Jiang2017, Barkoutsos2019}. There is evidence suggesting that QAOA may provide a significant quantum advantage over classical algorithms ~\cite{Guerreschi2019, Niu2019, Barkoutsos2019}, and that it is computationally universal~\cite{Morales2019universality, Lloyd2018}.
        
        Despite these advances, there are limitations and potential issues with this algorithm. The performance improves with the number of alternating layers in the ansatz, but the latter is limited by coherence times in existing and near-term quantum processors. Moreover, more layers implies more variational parameters, which introduces challenges for the classical optimizer~\cite{Wierichs2020avoiding}.  
        Furthermore, Ref.~\cite{Bravyi2019obstacles} points out that the locality and symmetry of QAOA can also severely limit its performance. These issues can be attributed to, or are at least exacerbated by, the form of the QAOA ansatz. In particular, short-depth QAOA is not really the digitized version of the adiabatic problem, but rather an ad hoc ansatz, and as a result should not be expected to perform optimally, or even well. A short-depth ansatz that is further tailored to a given combinatorial problem could therefore address the issues with the standard QAOA ansatz. However, identifying such an alternative is a highly nontrivial problem given the vast space of possible ans\"atze. 
        
        In this work, we propose an iterative version of QAOA termed Adaptive Derivative Assembled Problem Tailored - Quantum Approximate Optimization Algorithm (ADAPT-QAOA). Our algorithm grows the ansatz two operators at a time by using a gradient criterion to systematically select the QAOA mixer from a pre-defined operator pool. 
        While ADAPT-QAOA is general and can be applied to any optimization problem, we focus on Max-Cut  to quantify its performance. 
        We find that when entangling gates are included in the operator pool, there is a dramatically faster convergence compared to standard QAOA. Surprisingly, despite the introduction of entangling gates as mixers, these improvements come with a reduction in the numbers of both the optimization parameters and the CNOT gates by approximately 50\% each compared to standard QAOA. We summarize the details of our algorithm and its performance in Sec.~\ref{sec:ADAPT-QAOA} and
        provide evidence that this improved performance is related to the concept of shortcuts to adiabaticity \textcolor{blue}{\cite{demirplak2003adiabatic,Berry2009,Guery-Odelin2019, hegade2021shortcuts}} in Sec.~\ref{sec:STA}. Finally, we conclude in Sec. ~\ref{sec:conclusion} with a look ahead.
        
    \section{ADAPT-QAOA}\label{sec:ADAPT-QAOA}
    
    \subsection{Framework}
    
       In QAOA~\cite{Farhi2014, Farhi2014design}, the variational ansatz consists of $p$ layers, each containing the cost Hamiltonian $H_{C}$ and a mixer, $H_M$:
        \begin{equation}
        \ket{\psi_p(\vec\gamma,\vec\beta)} = \left(\prod_{k=1}^p\left[e^{-i H_{M} \beta_k}e^{-i H_{C} \gamma_k}\right]\right)\ket{\psi_{\rm ref}},
        \label{eq:original_ansatz}
        \end{equation}
        where  $\ket{\psi_{\rm ref}}=\ket{+}^{\otimes n}$, $n$ is the number of qubits, and $\vec\gamma$ and $\vec\beta$ are sets of variational parameters. If these parameters are chosen such that $\bra{\psi_p(\vec\gamma,\vec\beta)}H_C\ket{\psi_p(\vec\gamma,\vec\beta)}$ is minimized, then the resulting energy and state provide an approximate solution to the optimization problem encoded in $H_C$. The accuracy of the result and the efficiency with which it can be obtained depend sensitively on $H_M$. In the standard QAOA ansatz, the mixer is chosen to be a single-qubit $X$ rotation applied to all qubits. 
       A few papers have suggested modifications to the standard QAOA ansatz for specific problems and hardware architectures \cite{Farhi2017Quantum,Hadfield2019,ZhihuiPRA2020}. 
        These interesting results reveal the potential advantages of the QAOA ansatz but do not provide a universal strategy for choosing mixers that works across a broad range of optimization problems.
        
        In this work, we replace the single, fixed mixer $H_M$ by a set of mixers $A_k$ that change from one layer to the next:
        \begin{equation}
        \ket{\psi_p(\vec\gamma,\vec\beta)} = \left(\prod_{k=1}^p\left[e^{-i A_k \beta_k}e^{-i H_{C} \gamma_k}\right]\right)\ket{\psi_{\rm ref}}.
        \label{eq:adapt_ansatz}
        \end{equation}
        We build up this ansatz iteratively, one layer at a time, in a way that is determined by $H_C$. This iterative process is inspired by the variational quantum eigensolver algorithm, ADAPT-VQE \cite{Grimsley2018,Tang2019}. It can be summarized by three basic steps: First, define the operator set $\{A_j\}$ (called the ``mixer pool'', and where $A_j=A_j^\dagger$) and select a suitable reference state to be the initial state: $\ket{\psi^{(0)}}=\ket{\psi_{\rm ref}}$. Here, we choose $\ket{\psi_{\rm ref}}=\ket{+}^{\otimes n}$ as in the standard QAOA. We will return shortly to the question of how to choose the pool. Second, prepare the current ansatz $\ket{\psi^{(k-1)}}$ on the quantum processor and measure the energy gradient with respect to the pool, the $j$th component of which is given by  $-i\bra{\psi^{(k-1)}}e^{iH_C\gamma_{k}}[H_C,A_j]e^{-iH_C\gamma_{k}}\ket{\psi^{(k-1)}}$, where the new variational parameter $\gamma_{k}$ is set to a predefined value $\gamma_0$. 
        For the measurement, we can decompose the commutator into linear combinations of Pauli strings and measure the expectation values of the strings using general variational quantum algorithm methods~\cite{Google2019Measurement}.
        If the norm of the gradient is below a predefined threshold, then the algorithm stops, and the current state and energy estimate approximate the desired solution. If the gradient threshold is not met, modify the ansatz by adding the operator, $A_{\rm max}^{(k)}$, associated with the largest component of the gradient: $\ket{\psi^{(k)}}=e^{-iA^{(k)}_{\rm max}\beta_{k}}e^{-iH_C\gamma_{k}}\ket{\psi^{(k-1)}}$, where $\beta_{k}$ is a new variational parameter. 
        Third, optimize all parameters currently in the ansatz, $\beta_m$, $\gamma_m$, $m=1,...,k$, such that $\expval{H_C}{\psi^{(k)}}$ is minimized, and return to the second step. This algorithm, which we call ADAPT-QAOA, lies somewhere between standard QAOA and ADAPT-VQE in the sense that it possesses the alternating-operator structure of QAOA but enjoys additional flexibility by allowing the mixers to vary over the course of the iterative construction.
        
  \subsection{Operator Pool}

      \begin{figure*}
        \includegraphics[width=\linewidth]{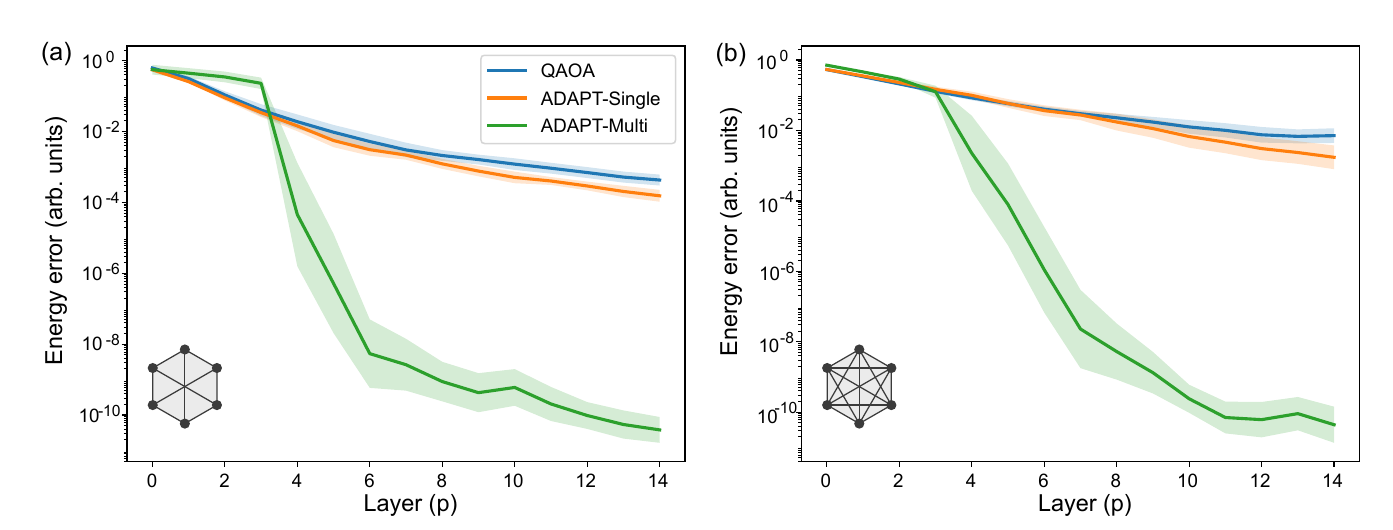}
        \caption{Comparison of the performance of standard QAOA (blue) with ADAPT-QAOA for the single-qubit (orange) and multi-qubit (green) pools. The algorithms are run on the Max-Cut problem for the regular graphs shown in the figure, which have $n{=}6$ vertices and are of degree $D{=}3$ (a) and $D{=}5$ (b). The energy error (the difference between the energy estimate obtained by the algorithm and the exact ground state energy of $H_C$) is shown as a function of the number of layers in the ansatz. Results are shown for 20 different instances of edge weights, which are randomly sampled from the uniform distribution $U(0,1)$. The shaded regions indicate $95\%$ confidence intervals.}
        \label{fig:performance}
    \end{figure*}
    
        The first step in running this algorithm is to define the mixer pool. Define $Q$ to be the set of qubits. The pool corresponding to the standard QAOA contains only one operator, $P_\text{QAOA} = \left\{ \sum_{i \in Q}X_i \right\}$. Here, we introduce two qualitatively different pools: one consisting entirely of single-qubit mixers, and one with both single-qubit and multi-qubit entangling gates: $P_\text{single} = \cup_{i \in Q} \left\{ X_i, Y_i\right\} \cup \left\{ \sum_{i \in Q}Y_i \right\} \cup P_\text{QAOA}$, $P_\text{multi} = \cup_{i,j \in Q \times Q} \left\{ B_i C_j | B_i, C_j \in \left\{ X, Y, Z \right\} \right\} \cup P_\text{single}$. Because $P_\text{QAOA} \subset P_\text{single} \subset P_\text{multi}$, we expect that $P_\text{multi}$ will provide the best performance. The QAOA, single-qubit, and multi-qubit pools have $O(1)$, $O(n)$, and $O(n^2)$ elements, respectively. 
       
        \begin{figure}
           \centering
           \includegraphics[width=\hsize,clip]{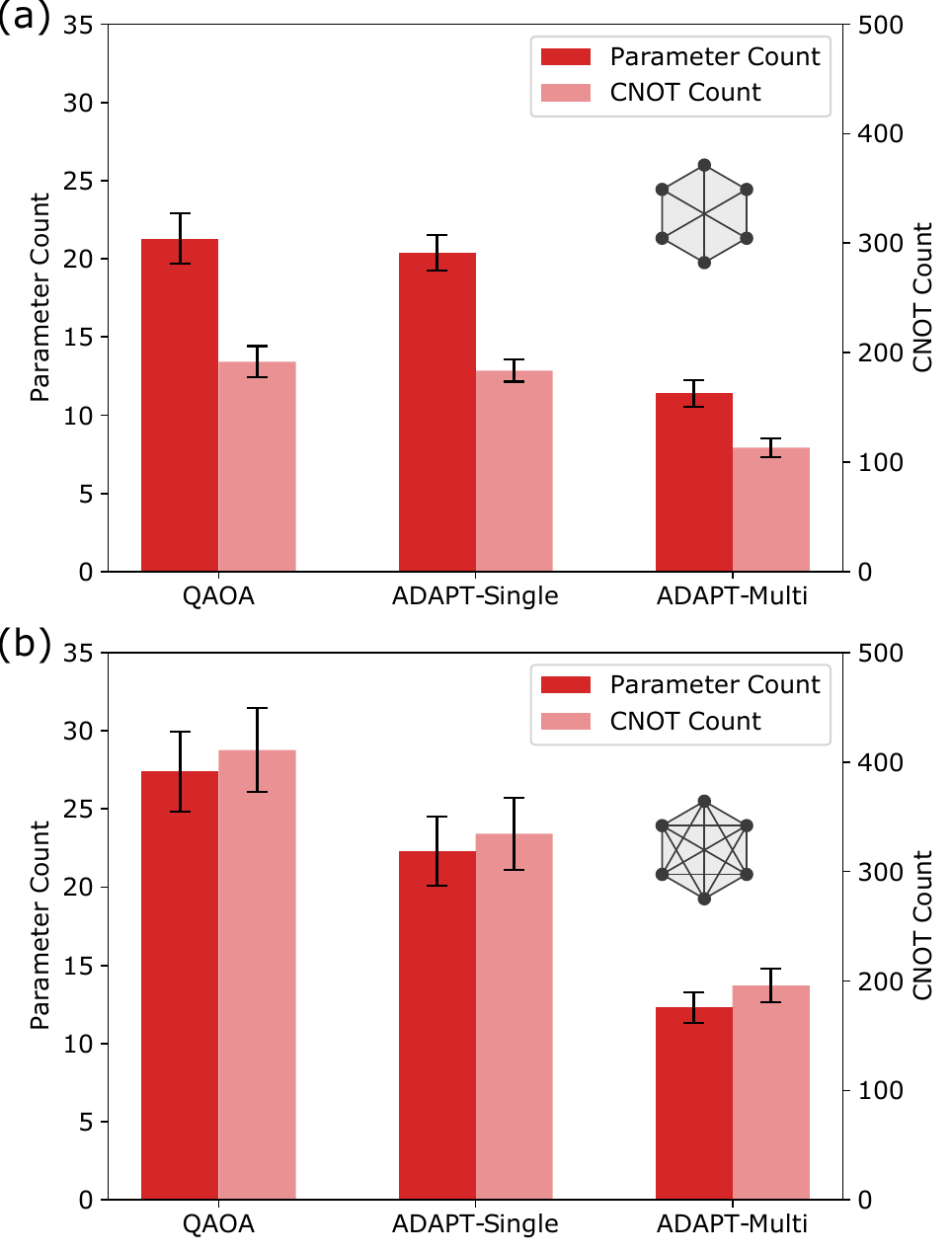}
           \caption{Resource comparison of the standard QAOA, ADAPT-QAOA with the single-qubit mixer pool, and ADAPT-QAOA with the multi-qubit mixer pool for the Max-Cut problem on regular graphs with $n{=}6$ vertices and random edge weights. Panels (a) and (b) show the comparison for graphs of degree $D{=}3$ and $D{=}5$, respectively. For all cases except the standard QAOA applied to $D=5$ graphs, we count the number of parameters and CNOTs needed to reach an energy error of $\delta E = 10^{-3}$. As standard QAOA for $D = 5$ graphs never reaches this error threshold, we instead count the CNOT gates and parameters at the end of the simulation (15 layers). The dark (light) red bars show variational parameter (CNOT gate) counts. The error bars show variances obtained by sampling over 20 different instances of edge weights.}
           \label{fig:rescource_comparison}
       \end{figure}
       
        Max-Cut is a classic (NP-hard) quadratic unconstrained binary optimization problem, and it can be used to solve other optimization problems. Thus, it is a useful benchmarking problem for QAOA and has been used as such in prior works~\cite{Farhi2014design,Crooks2018, Shaydulin2019}. It is defined as follows: Given a graph $G = (V, E)$, with weight $w_{i,j}$ for edge $(i,j)$, find a cut $S \subseteq V$ such that $S \cup \bar{S} = V$, and $\sum_{i\epsilon S, j \epsilon \bar{S}, i,j \epsilon E}w_{i,j}$ is maximized. This problem can be encoded in the Ising Hamiltonian 
        \begin{equation}
                H_{C} =-\frac{1}{2}\sum_{i,j}w_{i,j}(I-Z_{i}Z_{j}),\label{eq:isingham}
        \end{equation}
        where the couplings are given by the edge weights.
        Each classical state (i.e., tensor product of $Z$ eigenstates) represents a possible cut. $H_C$ counts the sum of the weights of the edges connecting one subgraph to the other, and its ground state corresponds to the maximum cut. $H_C$ has a $Z_2$ symmetry generated by $F=\otimes_i X_i$. Only the $A_j$ that commute with $F$ have a nonzero gradient (see Appendix~\ref{app:ising_sym}), so we retain only these Pauli strings (which have an even number of $Y$ or $Z$ operators) in our mixer pool. 
    \subsection{Performance and Resource Comparison}
        We use the Max-Cut problem on regular graphs with $n{=}6$ vertices and degrees $D{=}3$ and $D{=}5$ to benchmark the performance of ADAPT-QAOA. For each type of graph, we analyze 20 instances of random edge weights, which are drawn from the uniform distribution $U(0,1)$~\cite{Li2020AAS}. We use  Nelder-Mead for the optimization of the variational parameters $\vec{\beta}$ and $\vec{\gamma}$. The gradients used to select new operators are sensitive to the initial values for $\vec\gamma$. 
        It is natural to initialize these parameters at $\gamma_0=0$ to avoid biasing the optimization. However, as we show in Appendix~\ref{app:first_layer},
        $\gamma_0 = 0$ is a critical point of the cost function~\cite{shaydulin2019multistart}. Moreover, the  minimum of the energy in the first layer of ADAPT-QAOA never occurs at $\gamma_0 = 0$. Therefore, we shift the initial value $\gamma_0$ slightly away from zero ($\gamma_0=0.01$) to avoid these issues.
        
In Fig.~\ref{fig:performance} we show the error as a function of the number of ansatz layers for the standard QAOA and for ADAPT-QAOA using single-qubit and multi-qubit mixer pools. For both 3- and 5-regular graphs, we find that using the single-qubit mixer pool provides a modest improvement over standard QAOA. On the other hand, the multi-qubit pool performs dramatically better, leading to a rapid convergence to the exact solution after only $\sim$3 layers. We also find that for the degree-5 graphs, standard QAOA and ADAPT-QAOA with single-qubit mixers converge slower than the degree-3 case, whereas the performance of ADAPT-QAOA with the multi-qubit operator pool remains approximately the same. 
Note that the particular form of the two-qubit operators in the pool was chosen for its simplicity. In general, one can choose a hardware-tailored operator pool, in the spirit of Ref.~\cite{Google2020QAOA}. In Appendix~\ref{app:scalability}, we show similar results for $n=8$ and $n=10$ graphs of degree $D=2$, where ADAPT-QAOA with the multi-qubit pool substantially outperforms the standard QAOA again. Going to larger values of $D$ or $n$ is made challenging by a sharp increase in the number of layers needed to reach convergence, as reported for standard QAOA in Ref.~\cite{Akshay2020reachability}.
       
It is interesting to ask how much the ADAPT-QAOA ans\"atze differ from the standard QAOA ansatz. We find that when the single-qubit mixer pool is used, the single-qubit operators $X_i$ are chosen instead of the standard mixer approximately 36.6\% of the time for $n{=}6, D{=}3$ graphs and 25\% of the time for $n{=}6, D{=}5$ graphs. For the multi-qubit mixer pool, the algorithm chooses operators other than the standard mixer approximately 75\% of the time for $n{=}6, D{=}3$ graphs and 80\% of the time for $n{=}6, D{=}5$ graphs (see Appendix~\ref{app:mixers}). This trend supports the intuitive idea that a more connected graph requires more entanglement for a rapid convergence to the solution.

    A crucial question, especially for near-term platforms, is how the different mixer pools compare with respect to resource overhead. Fig.~\ref{fig:rescource_comparison} shows the number of CNOTs and number of parameters for the three algorithms. The CNOT counts are determined by decomposing each two-qubit operator into two CNOT gates and one or two single-qubit gates. Surprisingly, we find that both the standard QAOA and the single-qubit mixer ans\"atze in fact have more CNOTs compared to that constructed from the entangling multi-qubit mixer pool. Moreover, on average, the standard QAOA algorithm uses more parameters and CNOTs to reach the same convergence threshold than either version of ADAPT-QAOA. About half as many CNOTs are required for the ADAPT-QAOA multi-qubit pool case, despite the fact that the mixers in the multi-qubit pool themselves introduce additional CNOT gates on top of those coming from $H_C$. Ref.~\cite{Farhi2017Quantum} proposed using a restricted form of entangling gates in the ansatz to obtain better performance in combinatorial problems at the cost of introducing more variational parameters. In contrast, ADAPT-QAOA provides a systematic way to both improve performance and reduce the number of parameters and CNOTs.

\section{Shortcuts to adiabaticity} \label{sec:STA}

One may wonder whether there is a physically intuitive way to understand the strikingly better performance of ADAPT-QAOA. Considering that the standard QAOA ansatz has a structure dictated by the adiabatic theorem, a possible explanation is that the ADAPT algorithm is related to \textit{shortcuts to adiabaticity} (STA). STA, also known as counterdiabatic or transitionless driving, was introduced for quantum systems by Demirplak and Rice~\cite{demirplak2003adiabatic} and later, independently, by Berry~\cite{Berry2009, Guery-Odelin2019}. STA has also been explored in the classical context~\cite{Jarzynski2013,deng2013boosting}, including a recent application in biology~\cite{Iram2020}. The idea is that if we want to drive a system such that it remains in the instantaneous ground state at all times, then by adding a certain term ${\cal H}_{CD}$ to the Hamiltonian, we can achieve this \textit{without} paying the price of a slow evolution. Although the instantaneous eigenstates of the original Hamiltonian only solve the time-dependent Schr\"odinger equation in the adiabatic limit, they become exact solutions when the Hamiltonian is updated to include ${\cal H}_{CD}$. The advantage of STA is that the evolution can be achieved nonadiabatically. Below, we provide evidence that ADAPT-QAOA is indeed related to STA, a likely explanation for why it converges to the solution much faster than its adiabatic counterpart, the standard QAOA. Before we present this evidence, we must first explain how ${\cal H}_{CD}$ can be constructed using the concept of adiabatic gauge potentials.

\subsection{Approximate adiabatic gauge potentials}

Here we briefly review the mathematical machinery of STA and adiabatic gauge potentials~\cite{Kolodrubetz2017,Sels2017,Claeys2019}.
Let $\ket{\psi}$ be a state evolving under $\mathcal{H}(\theta(t))$, $i\partial_t\ket{\psi}=\mathcal{H}(\theta(t))\ket{\psi}$, where $\theta$ is a continuous variable that parameterizes the Hamiltonian. A unitary transformation $U(\theta(t))$ can be applied to move the Hamiltonian $\mathcal{H}(\theta(t))$ from the initial basis to its instantaneous eigenbasis, where $\tilde{\mathcal{H}}(\theta)=U^{\dagger}(\theta)\mathcal{H}(\theta)U(\theta)$ is diagonal at all times. The Schr\"{o}dinger equation in the instantaneous eigenbasis is
$i\partial_t\ket*{\tilde{\psi}}=[\tilde{\mathcal{H}}-\dot{\theta}\tilde{\mathcal{A}}_{\theta}]\ket*{\tilde{\psi}}$, where $\ket*{\tilde{\psi}}=U^{\dagger}\ket{\psi}$, $\dot{\theta}=d\theta/dt$, and $\tilde{\mathcal{A}}_{\theta}=iU^{\dagger}\partial_{\theta}U$ is the adiabatic gauge potential in the rotated frame. It is evident that the term $-\dot{\theta}\tilde{\mathcal{A}}_{\theta}$ drives transitions between the energy levels of the original Hamiltonian $\mathcal{H}$. Therefore, one can add the counterdiabatic term $\mathcal{H}_{CD}=\dot{\theta}\mathcal{A}_{\theta}$ to $\mathcal{H}(\theta)$, with $\mathcal{A}_{\theta}=U\tilde{\mathcal{A}}_{\theta}U^{\dagger}$, to eliminate such transitions in the rotated frame. This is the core of transitionless driving protocols. 

Now, the matrix elements of the adiabatic gauge potential in the instantaneous eigenbasis are
\begin{equation}
    \begin{aligned}
    \bra{m(\theta)}\mathcal{A}_{\theta}\ket{n(\theta)}&= \bra{m(\theta)}U\tilde{\mathcal{A}}_{\theta}U^{\dagger}\ket{n(\theta)}\\
    &=i\bra{m(\theta)}\partial_{\theta}UU^{\dagger}\ket{n(\theta)}\\
    &=i\bra{m(\theta)}\ket{\partial_{\theta}n(\theta)},\\
    \end{aligned}
\end{equation}
where we used $\tilde{\mathcal{A}}_{\theta}=iU^{\dagger}\partial_{\theta}U$ and $\ket{n(\theta)}=U(\theta)\ket{n_0}$ with $\ket{n_0}$ being independent of $\theta$. Moreover, the adiabatic gauge potential $\mathcal{A}_{\theta}$ satisfies~\cite{Berry2009,Kolodrubetz2017}
\begin{equation}
   \bra{m} \mathcal{A}_{\theta}\ket{n}=i\bra{m}\ket{\partial_{\theta}n}=i\frac{\bra{m}\partial_{\theta}\mathcal{H}\ket{n}}{E_n-E_m},
\end{equation}
which is obtained by differentiating the eigenfunction $\mathcal{H}(\theta)\ket{n(\theta)}=E_n(\theta)\ket{n(\theta)}$ with respect to $\theta$. Note that increasing the size of the system can lead to divergent matrix elements due to exponentially small denominators ($E_n-E_m$). In this regard, Ref.~\cite{Claeys2019} proposes an approximate gauge potential
\begin{equation}\label{eq:approx_H_cd}
    \mathcal{A}_{\theta}^{(p)}=i\sum_{k=1}^p a_k[\mathcal{H},\partial_{\theta}\mathcal{H}]_{2k-1},
\end{equation}
where $[X,Y]_{k+1}=[X,[X,Y]]_k$ and $\{a_1,a_2,\ldots,a_p\}$ is a set of coefficients with $p$ being the order of the expansion. This set of coefficients is found by minimizing $\Tr[G^2(\mathcal{A}_{\theta}^{(p)})]$, where 
$G(\mathcal{A}_{\theta}^{(p)})=\partial_{\theta}\mathcal{H}-i[\mathcal{H},\mathcal{A}_{\theta}^{(p)}]$~\cite{Claeys2019}. In fact, $\Tr[G^2(\mathcal{X})]$ is minimized when $\mathcal{X}$ is equal to the exact adiabatic gauge potential $\mathcal{A}_{\theta}$~\cite{Sels2017,Kolodrubetz2017}. Using matrix calculus identities and properties of the trace, it is straightforward to show that
\begin{equation}
    \frac{\partial \Tr[G^2(\mathcal{X})]}{\partial\mathcal{X}}=2[\mathcal{H},i\partial_{\theta}\mathcal{H}-[\mathcal{X},\mathcal{H}]].
\end{equation}
Only adiabatic gauge potentials satisfy $[\mathcal{H},i\partial_{\theta}\mathcal{H}-[\mathcal{A}_{\theta},\mathcal{H}]]=0$. This is easily proven by differentiating $\tilde{\mathcal{H}}(\theta)=U^{\dagger}(\theta)\mathcal{H}(\theta)U(\theta)$ with respect to $\theta$,
\begin{equation}
    \partial_{\theta}\tilde{\mathcal{H}}=\partial_{\theta}U^{\dagger}UU^{\dagger}\mathcal{H}U+U^{\dagger}\partial_{\theta}\mathcal{H}U+U^{\dagger}\mathcal{H}UU^{\dagger}\partial_{\theta}U,
\end{equation}
and noting that $\tilde{\mathcal{A}}_{\theta}=-i\partial_{\theta}U^{\dagger}U=iU^{\dagger}\partial_{\theta}U$ and $\mathcal{H}=\sum_n E_n(\theta)\ket{n(\theta)}\bra{n(\theta)}$, then
\begin{equation}
    [\mathcal{A}_{\theta},\mathcal{H}]=i(\partial_{\theta}\mathcal{H}-\sum_n\partial_{\theta}E_n(\theta)\ket{n(\theta)}\bra{n(\theta)}).
\end{equation}
Given that $[\mathcal{H},\sum_n\partial_{\theta}E_n(\theta)\ket{n(\theta)}\bra{n(\theta)}]=0$, adiabatic gauge potentials clearly satisfy
\begin{equation}
    [\mathcal{H},i\partial_{\theta}\mathcal{H}-[\mathcal{A}_{\theta},\mathcal{H}]]=0.
\end{equation}

\subsection{Connection between ADAPT-QAOA and STA}

\begin{figure}
  \centering
 \includegraphics[width=\linewidth]{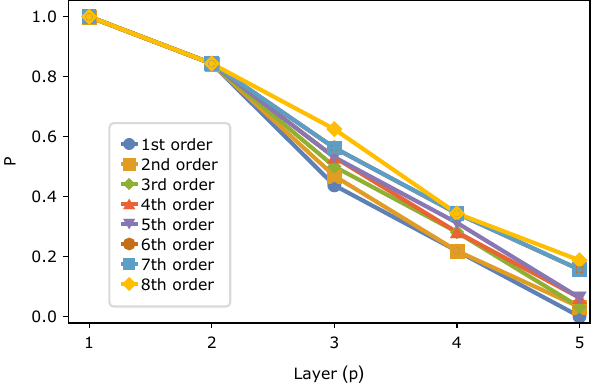}
\caption{Probability $P$ of the operator at layer $p$ of the ADAPT-QAOA ansatz to be among the Pauli strings with the largest coefficient in ${\cal H}_{CD}$ averaged over 32 graphs with $n=6,~D=3$. The different curves correspond to different orders of the approximation.}
    \label{fig:counterdiabatic_operators_vs_ADAPT}
    \end{figure}
   
To investigate the connection between ADAPT-QAOA and STA, we apply the above formalism using the Hamiltonian ${\cal H}=f(t)H_C+[1-f(t)]\sum_i^{n} X_i$ with $f(t)=t/T$ and, since there is no other continuous variable that parameterizes the Hamiltonian, we simply set $\theta=t$ in the equations above. $T$ is the duration of the evolution from the initial state $\ket{\psi_{ref}}=\ket{+}^{\otimes n}$ to the ground state of the cost Hamiltonian $H_C$, which is given by Eq.~\eqref{eq:isingham}. The counterdiabatic Hamiltonian ${\cal H}_{CD}$ is approximated using Eq.~\eqref{eq:approx_H_cd}, where $p$ is the order of the approximation.

As a concrete example, we study the Max-Cut problem on 32 instances of regular graphs ($n{=}6,~D{=}3$) with random edge weights. Fig.~\ref{fig:counterdiabatic_operators_vs_ADAPT} shows the probability that an operator in the ADAPT-QAOA ansatz is also one of the dominant operators in ${\cal H}_{CD}$. For each of the 32 cases, we define a set $\mathcal{O}_{CD}^{(i)}$ (with $i=1,\ldots,32$) comprised of the 5 operators with the largest coefficient in the time-averaged ${\cal H}_{CD}$ \footnote{We have seen in our simulations that only Pauli string operators are chosen by ADAPT-QAOA until deep into the layer number. Since we stop at layer 5, only Pauli strings are chosen in all 32 cases}. The probability $P$ in Fig.~\ref{fig:counterdiabatic_operators_vs_ADAPT} is constructed by taking the total number of times the mixer operator at layer $p$ is also an element of the corresponding set $\mathcal{O}_{CD}^{(i)}$ and dividing it by the total number of cases. In all cases, the mixer operator at the first layer is also an element of the set $\mathcal{O}_{CD}^{(i)}$. For higher layers, the probability of the mixer operator to be in $\mathcal{O}_{CD}^{(i)}$ is inversely proportional to the layer number. We attribute this to the fact that ${\cal H}_{CD}$ is computed for a specific mixer Hamiltonian ($\sum_i^{n} X_i$), while information about this choice does not enter into ADAPT-QAOA, which only relies on the initial state $\ket{+}^{\otimes n}$~\footnote{In principle, the mixer Hamiltonian in ${\cal H}(t)$ could be replaced by any other Hamiltonian that has $\ket{+}^{\otimes n}$ as the ground state, but this would modify the counterdiabatic Hamiltonian and any resemblance to the ADAPT-QAOA ansatz would be reduced or even lost. Interestingly, using the mixer Hamiltonian $\sum_i^{n} X_i$ involves the lowest possible energetic cost~\cite{Santos2015,zheng2016cost} of implementing the shortcut to adiabaticity.}. Interestingly, from Fig.~\ref{fig:counterdiabatic_operators_vs_ADAPT} we see that going to higher order in the ${\cal H}_{CD}$ approximation increases the probability of finding the mixers in the set $\mathcal{O}_{CD}^{(i)}$.
It therefore appears that ADAPT-QAOA finds the appropriate rotation axes in Hilbert space for faster convergence to the solution, and that these axes may in some sense be universal across all possible choices of ${\cal H}(t)$ that interpolate between the initial and target states. This suggests that STA can be used as a tool to construct operator pools for ADAPT-QAOA. 

\section{Conclusion}
\label{sec:conclusion}       
In conclusion, we introduced ADAPT-QAOA, a new optimization algorithm that grows the ansatz iteratively in a way that is naturally tailored to a given problem. We tested several instances of random diagonal Hamiltonians and found that ADAPT-QAOA always outperforms the standard QAOA.  Given its flexibility with the choice of mixer pool, the algorithm can be tailored to the native gates, connectivities, and experimental constraints of hardware. It would also be fruitful to employ ADAPT-QAOA for optimization problems that use higher-dimensional Hilbert spaces, such as graph coloring ~\cite{Hadfield2019,ZhihuiPRA2020}. Finally, more work into the connection to STA would be of both fundamental and practical interest.

\section*{ACKNOWLEDGEMENTS}
We thank Bryan T. Gard and Ada Warren for helpful discussions. S. E. E. acknowledges support from the US Department of Energy (Award No. DE-SC0019318). E.B. and N.J.M. acknowledge support from the US Department of Energy (Award No. DE-SC0019199).  

\appendix

\section{Ising symmetry and mixer pool operators}
\label{app:ising_sym}   

In this work, we focus on Ising Hamiltonians of the form
\setcounter{equation}{0}
\renewcommand\theequation{A.\arabic{equation}}

\begin{equation}
 H_{C} =-\frac{1}{2}\sum_{i,j}w_{i,j}(I-Z_{i}Z_{j}),
\label{eq:isingham_supp}
\end{equation}
which have a $Z_2$ symmetry associated with the operator $F=\otimes_i X_i$. Since $[F,H_C]=0$ and $F\ket{\psi_{\rm ref}}=\ket{\psi_{\rm ref}}$, we can rewrite the gradient in the first iteration as
\begin{align}
& \langle \psi_{\rm ref}| e^{iH_C\gamma_1} [H_C,A_j] e^{-iH_C\gamma_1} | \psi_{\rm ref}\rangle \nonumber \\
&=\langle \psi_{\rm ref}| e^{iH_C\gamma_1} F[H_C,A_j] Fe^{-iH_C\gamma_1} | \psi_{\rm ref}\rangle,
\label{eq:grad_flip}
\end{align}
where $A_j$ is an operator from the mixer pool.
However, we also know that $FA_j=\pm A_jF$ because $F$ and $A_j$ are Pauli strings (except when $A_j$ is the standard QAOA mixer $\sum_{i \in Q}X_i$ or $\sum_{i \in Q}Y_i$, but the former commutes and the latter anticommutes with $F$), so $F[H_C,A_j]F=\pm[H_C,A_j]$. Comparing this to Eq.~\eqref{eq:grad_flip}, we see that to have a non-zero gradient, we need $[F,A_j]=0$. This holds for all steps of the algorithm, because only operators that commute with $F$ appear in the ans\"atze, so a formula like Eq.~\eqref{eq:grad_flip} holds at every iteration. The $A_j$ that commute with $F$ are Pauli strings that have an even number of $Y$ or $Z$ operators, so we retain only these Pauli strings in our mixer pool. 

\section{First Layer of ADAPT-QAOA ansatz}\label{app:first_layer}
\setcounter{equation}{0}
\renewcommand\theequation{B.\arabic{equation}}

Here we analyze the ADAPT-QAOA cost function in the first layer, and the results show that the minimum of the energy in the first layer of ADAPT-QAOA never occurs at $\gamma=0$ for any operator included in the pool. We also show that $\gamma=0$ is a critical point of the cost function.

At level $p$ of ADAPT-QAOA, the cost function is
\begin{equation}
    E_p(\beta,\gamma)=\langle\psi^{(p-1)}|e^{-i\gamma H}e^{-i\beta M}He^{i\beta M}e^{i\gamma H}|\psi^{(p-1)}\rangle.
\end{equation}
Where $H$ is a linear combination of Pauli strings that are tensor products of the identity and $Z$. All the terms in $H$ commute with each other. $M$ is the mixer.
If the mixers $M$ are single-Pauli strings, we have
\begin{equation}
    e^{-i\beta M}He^{i\beta M}=H_c+\cos(2\beta)H_a-i\sin(2\beta)MH_a,\label{eq:identity}
\end{equation}
where $H_c$ is the part of $H$ that commutes with $M$, and $H_a$ is the part of $H$ that anticommutes with $M$. We then have

\begin{align}
    &e^{-i\gamma H}e^{-i\beta M}He^{i\beta M}e^{i\gamma H}\nonumber\\&
    =H_c+\cos(2\beta)H_a-i\sin(2\beta)e^{-i\gamma H}Me^{i\gamma H}H_a\nonumber\\&
    =H_c+\cos(2\beta)H_a-i\sin(2\beta)MH_ae^{2i\gamma H_a}.
\end{align}

We know that $E_p(\beta,\gamma)$ is periodic in $\beta$ with period $\pi$. Therefore, we can restrict $\beta$ to the range $\beta\in[-\pi/2,\pi/2]$ without loss of generality.
Let's define 
\begin{equation}
G(\gamma)\equiv -iMH_ae^{2i\gamma H_a}.\label{eq:defOfG}
\end{equation}
The cost function is then
\begin{equation}
    E_p(\beta,\gamma)=\langle H_c\rangle+\cos(2\beta)\langle H_a\rangle+\sin(2\beta)\langle G(\gamma)\rangle,
\end{equation}
where the expectation values are taken with respect to $|\psi^{(p-1)}\rangle$. Therefore,
\begin{eqnarray}
&&\frac{\partial E_p}{\partial\beta}=-2\sin(2\beta)\langle H_a\rangle+2\cos(2\beta)\langle G(\gamma)\rangle=0 \nonumber\\
&&\Rightarrow \tan(2\beta)=\frac{\langle G(\gamma)\rangle}{\langle H_a\rangle},\label{eq:dBetaEqualsZero}
\end{eqnarray}
and
\begin{eqnarray}
&&\frac{\partial E_p}{\partial\gamma}=\sin(2\beta)\langle G'(\gamma)\rangle=0 \nonumber \\
&&\Rightarrow \beta=0,\pm\pi/2\quad\hbox{or}\quad \langle G'(\gamma)\rangle=0.\label{eq:dGammaEqualsZero}
\end{eqnarray}

For the first layer, $p=1$, the state is $|\psi^{(0)}\rangle=|+\rangle^{\otimes n}$, and so $\langle H_a\rangle=0$. From Eq.~(\ref{eq:dBetaEqualsZero}), we see that $\beta=\pm\pi/4$ assuming $\vev{G(\gamma)}\ne0$. Eq.~(\ref{eq:dGammaEqualsZero}) then requires $\langle G'(\gamma)\rangle=0$. Using Eq.~\eqref{eq:defOfG}, for $p=1$ this is:

\begin{eqnarray}
\vev{G'(\gamma)} &&=2\vev{MH_a^2e^{2i\gamma H_a}}\nonumber \\ &&=2\vev{H_a^2e^{2i\gamma H_a}}=0.\label{eq:Gprime}
\end{eqnarray}

Notice that $\gamma=0$ is {\it not} a solution of this equation because $\vev{H_a^2}>0$, which follows from the fact that this is the norm of a nonzero state, $H_a|+\rangle^{\otimes n}$. Numerics are needed to determine if there is a nonzero value of $\gamma$ that satisfies $\vev{G'(\gamma)}=0$.

On the other hand, if $\vev{G(\gamma)}=0$, then there is no constraint on $\beta$ from $\frac{\partial E_p}{\partial\beta}=0$. Notice that this happens when $\gamma=0$ since $\vev{G(0)}=-i\vev{MH_a}=0$, which is true for regular graphs where $H_a$ is a sum of terms like $Z_jZ_k$, and $M$ is a Pauli string that commutes with $F=\otimes_\ell X_\ell$. In this case, we must have $\beta=0$ or $\pm\pi/2$. There could be other values of $\gamma$ that satisfy $\vev{G(\gamma)}=0$, but it seems unlikely that both this condition and $\vev{G'(\gamma)}=0$ can be satisfied simultaneously, so it is probably still true that $\beta$ must be 0 or $\pm\pi/2$ in this case.

In summary, there are two classes of possible solutions for $p=1$: (i) $\beta=\pm\pi/4$ and $\gamma=\gamma^*\ne0$ where $\vev{G'(\gamma^*)}=0$; (ii) $\gamma=\gamma^*$ where $\vev{G(\gamma^*)}=0$ and either $\vev{G'(\gamma^*)}=0$ or $\beta=0,\pm\pi/2$.
To clarify whether these solutions are maxima or minima, we calculated the second derivatives:
\begin{align}
\frac{\partial^2 E_p}{\partial\beta^2}&=-4\cos(2\beta)\vev{H_a}-4\sin(2\beta)\vev{G(\gamma)},\\
\frac{\partial^2 E_p}{\partial\gamma^2}&=\sin(2\beta)\vev{G''(\gamma)},\\
\frac{\partial^2 E_p}{\partial\beta\partial\gamma}&=2\cos(2\beta)\vev{G'(\gamma)}.
\end{align}
Let's consider the class (i) extrema first.
In this case
\begin{eqnarray}
\frac{\partial^2 E_p}{\partial\beta^2}\bigg|_{\beta=\pm\pi/4,\gamma=\gamma^*}=\mp4\vev{G(\gamma^*)},
\end{eqnarray}
and the Hessian is
\begin{eqnarray}
&& \left\{\frac{\partial^2 E_p}{\partial\beta^2}\frac{\partial^2 E_p}{\partial\gamma^2}-\left(\frac{\partial^2 E_p}{\partial\beta\partial\gamma}\right)^2\right\}\bigg|_{\beta=\pm\pi/4,\gamma=\gamma^*} \nonumber \\
&& =-4\vev{G(\gamma^*)}\vev{G''(\gamma^*)}.
\end{eqnarray}
Thus, we need $\vev{G(\gamma^*)}\vev{G''(\gamma^*)}<0$, in which case $\beta=\pi/4$ is a maximum and $\beta=-\pi/4$ is a minimum if $\vev{G(\gamma^*)}>0$, while $\beta=-\pi/4$ is a maximum and $\beta=\pi/4$ is a minimum if $\vev{G(\gamma^*)}<0$.

For type (ii) extrema, we have
\begin{eqnarray}
\frac{\partial^2 E_p}{\partial\beta^2}\bigg|_{\beta=0,\pm\pi/2,\gamma=\gamma^*}=\pm4\vev{H_a}=0,
\end{eqnarray}
and the Hessian is
\begin{eqnarray}
&&\left\{\frac{\partial^2 E_p}{\partial\beta^2}\frac{\partial^2 E_p}{\partial\gamma^2}-\left(\frac{\partial^2 E_p}{\partial\beta\partial\gamma}\right)^2\right\}\bigg|_{\beta=0,\pm\pi/2,\gamma=\gamma^*} \nonumber \\
&&=-4\vev{G'(\gamma^*)}^2,
\end{eqnarray}
these extrema are saddle points. Notice that $\gamma=0$ is one of these saddle points.

We can see that the minimum of the energy in the first layer of ADAPT-QAOA never occurs at $\gamma=0$, and the only minima occur at $\beta=\pm\pi/4$ and $\gamma=\gamma^*\ne0$, which is consistent with our numerical simulation results. The fact that $\gamma=0$ is a saddle point (at least in the first layer) makes this a problematic choice for the initial value of $\gamma$ in the classical optimization subroutine of ADAPT-QAOA. For this reason, we instead choose the initial value, $\gamma_0$, to be a small nonzero value, e.g., $\gamma_0=0.01$. Similar performance is achieved for any value of $\gamma_0\lesssim 0.1$. Keeping $\gamma_0$ close to zero is preferable to avoid biasing the optimization.

\section{Scalability of ADAPT-QAOA}\label{app:scalability}

\begin{figure}[!h]
    \includegraphics[width=0.99\linewidth]{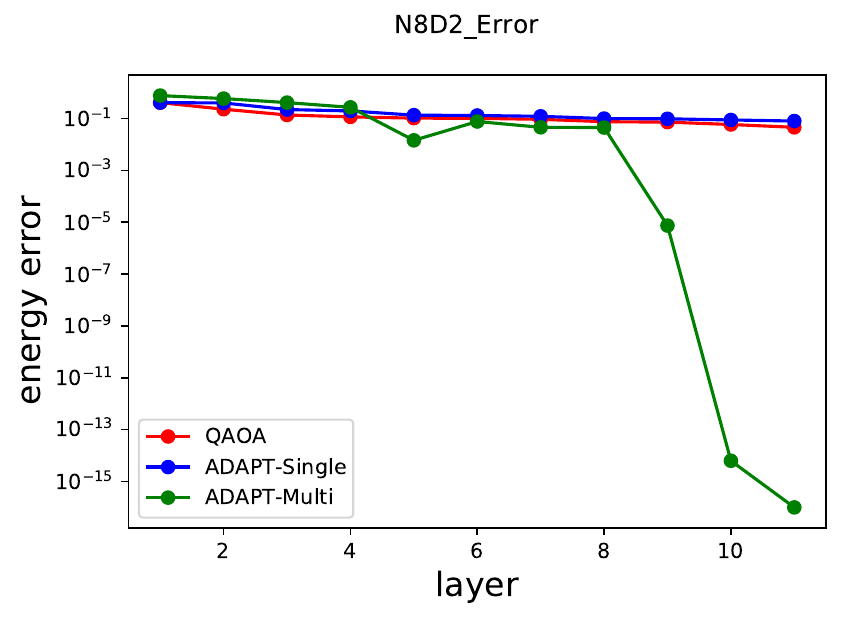}
    \caption{The performance of three algorithms for the Max-Cut problem on a $n = 8$ and $D = 2$ graph with randomly chosen edge weights.}
    \label{fig:N8D2}
\end{figure}

\begin{figure}[!h]
    \includegraphics[width=0.99\linewidth]{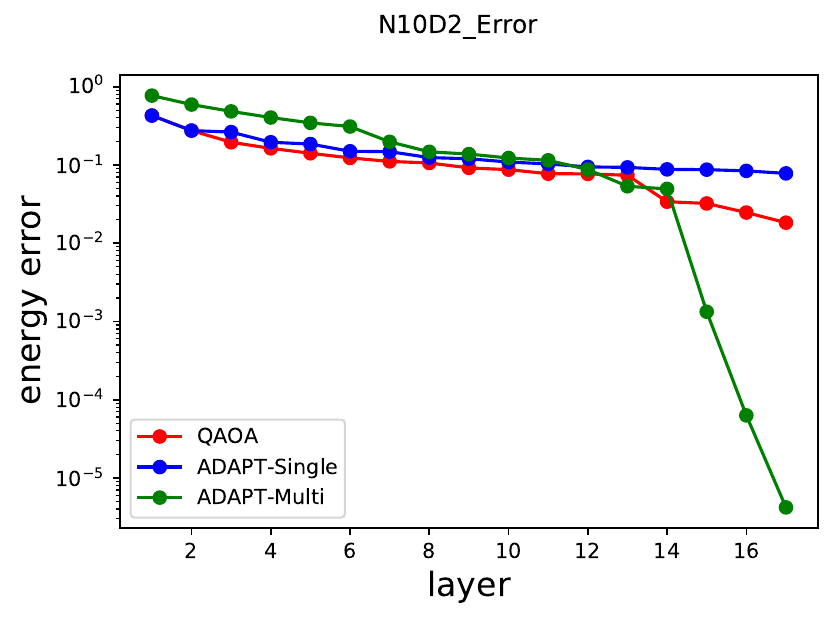}
    \caption{The performance of three algorithms for the Max-Cut problem on a $n = 10$ and $D = 2$ graph with randomly chosen edge weights.}
    \label{fig:N10D2}
\end{figure}

Here, we investigate the scalability of our algorithm. We perform simulations for degree $D = 2$ graphs with $n=8$ and $n=10$ qubits, with results shown in Figs.~\ref{fig:N8D2} and \ref{fig:N10D2}. Clearly, even for larger system sizes and less connected graphs where $D / n < 1 / 2$, our algorithm exhibits substantially better performance compared to standard QAOA.

\section{Role of entangling mixers versus entangling gates}\label{app:mixers}
\setcounter{equation}{0}
\renewcommand\theequation{C.\arabic{equation}}

\begin{figure*}[!ht]
    \includegraphics[width=0.90\linewidth]{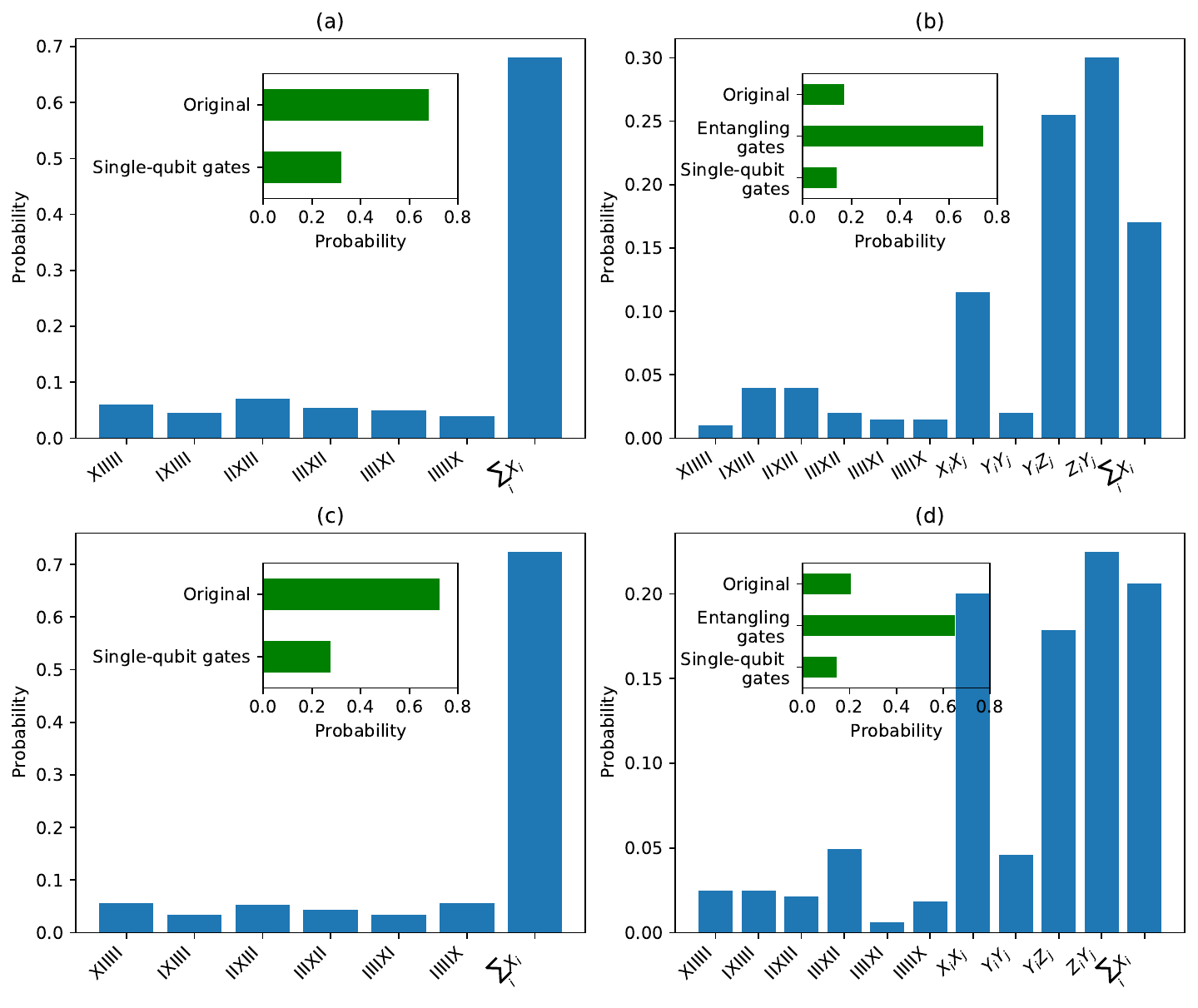}
    \caption{Probability of operators picked by the original QAOA, ADAPT-QAOA with the single-qubit mixer and ADAPT-QAOA with multi-qubit pool for the Max-Cut problem on regular graphs with $n$=6 vertices with degree $D$=3 (a)(b) and $D$=5 (c)(d) with random edge weights sampled from a uniform distribution $U(0,1)$. The blue bars show the probability of each particular operator used for ansatz, and green bars show the probability of the original mixer, sum over all single-qubit gates and sum over all entangling gates used in ansatz. The results from 20 instances of random edge weights. }
    \label{fig:operator_frequency}
\end{figure*}

\begin{figure}[!ht]
    \includegraphics[width=0.95\linewidth]{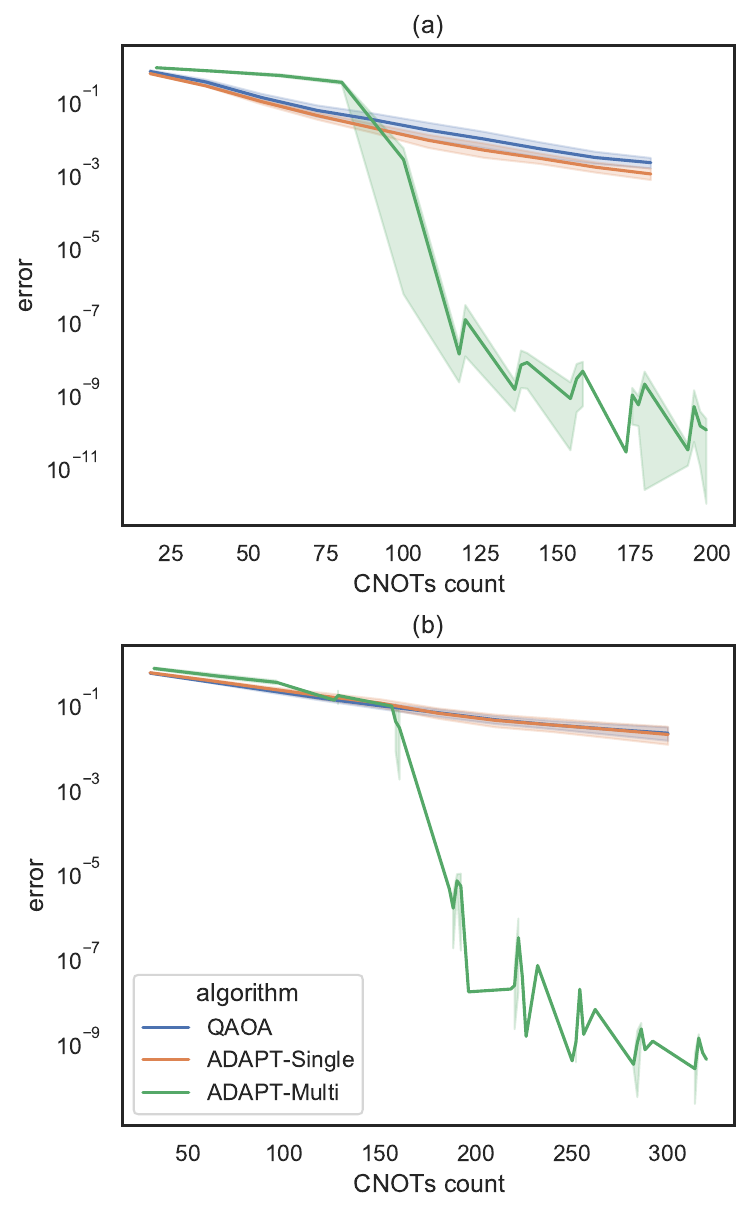}
    \caption{Comparison of the performance of the original QAOA algorithm (blue) with the ADAPT-QAOA algorithm for the single-qubit (orange) and multi-qubit (green) pools run for the Max-Cut problem on graphs with $n=6$ vertices with degree $D=3$ (a) and $D=5$ (b). The energy error is shown as a function of the number of total CNOTs used in the ansatz.}
    \label{fig:performance_CNOTs}
\end{figure}

In Fig.~\ref{fig:rescource_comparison}, we compared the resources used by three different algorithms for the Max-Cut problem on regular graphs. These resources include the number of CNOT gates and the number of optimization parameters. From the comparison we can see that including entangling mixers in the ansatz produces a dramatically faster convergence to the exact solution compared to the original QAOA. Surprisingly, despite the inclusion of these entangling mixers, the improvement in convergence comes with a simultaneous reduction in both the number of entangling gates in the compiled ansatz.

To further investigate the role of entangling mixers, we consider the $n=6,~D=3$ (in Fig.~\ref{fig:operator_frequency} (a), (b)) and $n=6,~D=5$ graphs (in Fig.~\ref{fig:operator_frequency} (c), (d)) and show the probability for an operator to be picked by the original QAOA and by ADAPT-QAOA with a single-qubit mixer pool in Fig.~S1. Similar results for ADAPT-QAOA with a multi-qubit mixer pool are also shown in Fig.~\ref{fig:operator_frequency}(b),(d). We find that when only the single-qubit mixer pool is used, the single-qubit operators $X_i$ are chosen instead of the original mixer approximately $25\%$ of the time. For the multi-qubit mixer pool, the algorithm chooses two-qubit entangling operators approximately $70\%$ of the time. Clearly, entangling mixers play a central role in the improved performance of ADAPT-QAOA.

Additionally, to understand the importance of CNOT gates in the compiled ansatz, in Fig.~\ref{fig:performance_CNOTs} we show the error for the solution determined by each algorithm as a function of the number of CNOTs used in the ansatz for both $n=6,~D=3$ and $n=6,~D=5$ graphs. Based on the figure, we can see that ADAPT-QAOA with the multi-qubit mixer pool converges using a much smaller number of CNOTs compared to the original QAOA or to ADAPT-QAOA with the single-qubit mixer pool. On the other hand, the latter only provides a modest reduction in CNOT count compared to the original QAOA. Interestingly, we see that while entangling mixers are crucial for the dramatic improvement in the performance afforded by ADAPT-QAOA with the multi-qubit mixer pool, at the same time far fewer entangling gates are needed in the ansatz. Thus, the role of entanglement in this algorithm is rather subtle.


%


\end{document}